\documentclass[aps,showpacs,amsmath,amssymb,nofootinbib,preprintnumbers]{revtex4}
\usepackage{graphicx}
\usepackage{color}
\def \be  {\begin{equation}}
\def \ee  {\end{equation}}
\def \bea {\begin{eqnarray}}
\def \eea {\end{eqnarray}}

\begin{document}

\preprint{ECTP-2011-04}

\title{Thermodynamics of viscous Matter and Radiation in the Early universe}
\author{A.~Tawfik}
\email{a.tawfik@eng.mti.edu.eg}
\email{atawfik@cern.ch}
\affiliation{Egyptian Center for Theoretical Physics (ECTP), MTI University,
Cairo, Egypt}
\affiliation{Research Center for Einstein Physics, Freie-University Berlin, Berlin, Germany}

\author{H.~Magdy}
\affiliation{Egyptian Center for Theoretical Physics (ECTP), MTI University,
Cairo-Egypt}

\date{\today}

\begin{abstract}
Assuming that the background geometry is filled with a free gas consisting of matter and radiation and that no phase transitions are occurring in the early universe, we discuss the thermodynamics of this {\it closed} system using classical approaches. We find that essential cosmological quantities, such as the Hubble parameter $H$, scale factor $a$, and curvature parameter $k$, can be derived from this simple model. On one hand, it obeys the laws of thermodynamics entirely. On the other hand, the results are compatible with the Friedmann–Lemaitre–Robertson–Walker model and the Einstein field equations. The inclusion of finite bulk viscosity coefficient derives to important changes in all Of these cosmological quantities. The thermodynamics of the viscous universe is studied and a conservation law is found. Accordingly, our picture of the evolution of the early universe and its astrophysical consequences seems to be the subject of radical revision. We find that the parameter $k$, for instance, strongly depends on the thermodynamics of the background matter. The time scale at which a negative curvature might take place, depends on the relation between the matter content and the total energy. Using quantum and statistical approaches, we assume that the size of the universe is given by the volume occupied by one particle and one photon. Different types of interactions between matter and photon are taken into account. In this quantum treatment, expressions for $H$ and $a$ are also introduced. Therefore, the expansion of the universe turns out to be accessible.
\end{abstract}


\maketitle

\section{Introduction} \label{sec:1}
The equations of state (EoS) describing the matter and/or radiation filling the cosmological background geometry play an essential role in determining the time evolution of the universe. The more reliable EoS are included, the more realistic the time resultant evolution is. Such models are favoured because we—so far—have no observational evidence for the {\it real} time evolution, especially during the very early eras of the universe. The present work is an extension of a previous one \cite{Tawfik:2010ht}, where the cosmic background matter is conjectured to consist of two parts. The first part is a massive particle with mass $m$. The second part is given by an {\it absolute} space (frame of reference)(i.e.  that is {\it absolute} background) with mass $M$. Such a background is characterized by Newton's theory \cite{newton}, which relates the first and second laws to an {\it absolute} space. Rather than the mass and the gravitational field, the background itself possesses no other features. Seeking for completeness, we mention here that Mach reasoned that Newton's postulates, which are relative to the simplicity of the newtonian laws, are related to the large scale distribution of matter in the universe \cite{mach}. Mach's postulates have been implemented in the theory of special relativity \cite{sr}.

In the present work, we assume that the other matter components, like dark matter and dark energy, would not matter much during these early stages. On the other hand, we add a new component representing radiation. The present treatment avoids the inclusion of the relativistic mass of the photon. This might be the subject of a future work. Then, the cosmic background is to be determined by one particle with mass $m$ and a photon with energy $h\,\nu$, where $h$ is the Planck constant and $\nu$ is the frequency. We take into consideration two cases. In the first case, we disregard the interactions between particle and photon. In the second case, we consider interactions, especially, when the quantum nature is elaborated. Like hydrogravitational dynamics, the cosmological theory \cite{hgd} is based on hypotheses similar to the ones utilized in the present paper.

All phase transitions are disregarded and the background matter is assumed to be likely formed as a free gas (i.e.  nonviscous). We applied the laws of thermodynamics and the fundamentals of classical physics in order to derive expressions for basic cosmological quantities, such as the Hubble parameter $H(t)$, scale factor $a(t)$, and curvature parameter $k$. Then, we compared them with the Friedmann–Lemaitre–Robertson–Walker (FLRW) model and Einstein's field equations.

Unless it is explicitly stated, we assume natural units so that the selected universal physical constants are normalized to unity. We apply the standard cosmological model in order to gain global evidence supporting the FLRW model, although we disregard the relativistic and microscopic effects. The various forms of matter and radiation are conjectured to be homogeneously and isotropically distributed. We use non-relativistic arguments to give expressions for the thermodynamic quantities in the early universe, which obviously reproduce essential parts of the well-known FLRW model. We assume that the universe is in thermal equilibrium and therefore the interaction rates apparently exceed the universe expansion rate, which likely was being slowed down with an increase the comoving time $t$. Also, we assume that the expansion is adiabatic (i.e.  no entropy production or heat change takes place). Finally, we take into consideration two forms of the cosmic background matter. The first one is an ideal gaseous fluid, which is characterized by the lack of interactions and the constant internal energy. The second one is a viscous fluid, which is characterized by the long range correlations and the velocity gradient along the scale factor $a(t)$.

The present paper is organized as follows. The time evolutions of the energy density in nonviscous and viscous cosmology are discussed in sects. \ref{sec:2} and \ref{sec:3b}, respectively. The expansion rate itself in curved and flat universes is studied in section \ref{sec:3}. The quantum nature of the universe is introduced in section \ref{sec:4}. Finally, section \ref{sec:5} is devoted to discussion.

\section{Rate of Energy density in nonviscous cosmology}\label{sec:2}

Based on the model introduced in Ref. \cite{Tawfik:2010ht}, we first assume that all types of energies in the early universe are heat, $Q$, and the background geometry is filled with just one particle. Then
\bea \label{eq:dq}
dQ &=& 0 = dU +pdV,
\eea
where $U$ is the internal energy and $p$ is the pressure. The volume $V$ can be approximated to be proportional to $a^3$. It is obvious that (\ref{eq:dq}) is the first law of thermodynamics. The energy density $\rho=U/V$ likely decreases with the expansion of the universe; $d\rho=dU/V-U dV/V^2$. In comoving coordinates, $U$ is equivalent to the mass of the particle $m$. Then, from (\ref{eq:dq}), we get
\bea\label{eq:dro}
d\rho &=& -3 (p + \rho)\frac{da}{a}.
\eea
Dividing both sides by an infinitesimal time element $dt$ results in
\bea \label{eq:drot}
\dot \rho &=& -3 (p + \rho)\, H,
\eea
which is nothing but the equation of motion from the FLRW model at a vanishing cosmological constant, $\Lambda=0$, and curvature parameter, $k=0$. The time evolution of the energy density strongly depends on the thermodynamic quantities, $\rho$ and $p$, that is on the EoS of the matter and/or radiation occupying the background geometry. One dot means first derivative with respect to the comoving time $t$; $H$ is the Hubble parameter, which relates the velocity with the distance, $H=\dot a/a$.
When inserting one photon in the background geometry, then
\bea\label{eq:pg1}
\dot \rho &=& -3 \left((p+\rho) + \frac{2\,\pi}{a^3}\right)\; H.
\eea

The radiation-dominated phase is usually characterized by the equation of state $p=\rho/3$. Therefore (\ref{eq:drot}) leads to $\rho\propto a^{-4}$. In the matter-dominated phase, $p<<\rho$ and therefore $\rho\propto a^{-3}$, ($\rho\,\propto V^{-1}$). The energy density $\rho$ can be expressed in terms of the temperature $T$. Then, we can rephrase the proportionality in the radiation-dominated phase as $\rho\propto T V^{-1}$.


\section{Expansion Rate in Viscous Cosmology}\label{sec:3}

In previous sections, the meaning of adding a photon to the model \cite{Tawfik:2010ht} is introduced. It is supposed to represent the radiation filling the background geometry. Therefore, the cosmic background is now characterized by one particle with mass $m$ and one photon with frequency $\nu$. Also, it is conjectured that the photon completes one oscillation over the whole radius of the universe. Therefore, the photon's frequency $\nu=1$. Such an assumption fits well with the model \cite{Tawfik:2010ht}, where it has been conjectured that the expansion itself is determined by the distance covered by the particle with mass $m$. In other words, the size of the universe is given by the distance covered by the particle and simultaneously along which the photon is able to complete one cycle. These two components are located at a distance $a$ from some point in the universe. In the radial direction, the particle will have a kinetic energy $m\dot a^2/2$. The kinetic energy of the photon reads $h \nu$. In the opposite direction, both are affected by a gravitational force due to their masses $m_p$, $m_\gamma$ and the mass inside the sphere, which is given as $M=(4\pi/3)a^3 \rho$. The latter characterizes the mass of the {\it absolute} background. Then, the particle's {\it gravitational} potential energy is $-G M m_p/a$ and the photon's one is $-G M m_{\gamma}/a$, where $G$ is the newtonian gravitational constant. In natural units, $c=k_B=\hbar=G=1$, where $c$ and $k_B$ are speed of light and the Boltzmann constant, respectively. Then, the Planck constant $h=2\pi$. Therefore, the total energy reads
\bea \label{eq:tEnr1}
E&=&\frac{1}{2}\, m_p\, \dot a^2 - \frac{M\, m_p}{a}+2\,\pi - \frac{M\, m_\gamma}{a},
\eea
where the third term represents the photo's energy. For $m_p\gg{m_\gamma}$, (\ref{eq:tEnr1}) can be re-written as
\bea \label{eq:adot1}
\dot a^2 + k &=& \frac{8\pi}{3}\, \rho\, a^2.
\eea
This is nothing but the Friedman's first equation with the curvature parameter
\bea \label{eq:q2}
k &=& \frac{2}{m_p}\, \left(2\, \pi -E\right).
\eea
In Friedman's solution, $k$ can be vanishing  $\pm1$, referring to flat or positively or negatively curved universe, respectively~\cite{dverno}. It is straightforward to conclude that the value assigned to $k$ depends on the interplay between the positive and negative terms in (\ref{eq:k01}).

For a flat universe(i.e. $k=0$) it is very easy to find solutions for (\ref{eq:q2}). One solution leads to $m_p\rightarrow \infty$ (i.e. very heavy particle mass). The other solution relies on the photon's energy, where to the total energy is exclusively determined by the photon, $E=2\,\pi$. In order to omit $m_p$, (\ref{eq:q2}) can be rewritten as
\bea \label{eq:q2b}
k &=& -\dot{a}^2\, + \frac{2\,M}{a},
\eea
which has two solutions at $k=0$,
\bea
a &=& \left[\frac{3}{2}\left(c\pm \sqrt{2\, M}\; t\right)\right]^{2/3},
\eea
where $c$ is to be determined from the boundary conditions.

It is obvious that the quantities ${\cal O}(\pi) $ in (\ref{eq:adot1}) and (\ref{eq:q2}) are coefficients and the quantity $2\,\pi$ has the same dimension as the energy $E$ in natural units ( (\ref{eq:tEnr1}) and (\ref{eq:q2}), which refers to the energy of a photon that completes one cycle. For positively or negatively curved universe, the total energy is given by subtracting $m_p/2$ from and adding the same quantity to the photon's energy, respectively. In other words, the particle`s mass apparently determines the curvature of the early universe. If it is added (to the photon`s energy) it derives the universe to have closed curvature and vice versa.
\bea \label{eq:k01}
E &=& 2\pi\,\mp\, \frac{m_p}{2},
\eea
In general relatively, the curvature of space is related to the energy–momentum tensor. The relation is given by the Einstein's field equations \cite{eins1916}.
\bea \label{eq:eins1}
8 \pi\, \textbf{T}_{\mu\,\nu} &=&  \textbf{R}_{\mu\,\nu} -\frac{\textbf{g}_{\mu\,\nu}}{2} R,
\eea
where $\textbf{R}_{\mu\,\nu}$ and $R$ are the Ricci curvature tensor and scalar, respectively, and $\textbf{g}_{\mu\,\nu}$ is the metric tensor. Expression (\ref{eq:eins1}) relates the matter (energy) content to the universe`s curvature. To keep matching the assumptions of the present work, the cosmological constant is assumed to vanish in (\ref{eq:eins1}). Equation (\ref{eq:k01}), which is valid for $k=\pm 1$, leads to the conclusion that the curvature of the universe is positive, when the particle`s mass is subtracted. When the particle`s mass is added, then the universe`s curvature becomes negative. Obviously, such a result does not rely on a theory. The present work is merely designed as an effective model offering hints on the real evolution of the early universe.

According to recent heavy-ion collision experiments \cite{rhic1} and lattice QCD simulations \cite{nakam}, matter under extreme conditions (very high temperature and/or pressure) seems not to be, as we used to assume over the last three decades, an {\it ideal} gas, in which no collisions take place. It has been found that such matter is likely fluid, that is strongly correlated matter with finite heat conductivity and transport properties, especially finite viscosity coefficients (bulk and shear)~\cite{finiteta1}. Therefore, it is appropriate to apply this assumption on the background geometry in the early universe. This is the motivation of the present extension. For simplicity, we assume that the shear viscosity is almost negligible. Therefore, we assume that the cosmic background geometry should not necessarily be filled with an ideal free gas. In previous work~\cite{taw1,taw2,taw3,taw4,taw5a,taw6a}, we introduced models in which we included finite bulk viscosity coefficients. The analytical solutions of such models are nontrivial \cite{taw1,taw2,taw3,taw4,taw5a,taw6a}. In the present work, we try to introduce an approach for the viscous cosmology using this simple model, in which we just utilize classical approaches. As will be shown below, the classical approach seems to work perfectly in nonviscous background matter. It is in order now to check the influences of viscous fluid on the cosmological evolution. The simplicity of these approaches does not enhance the validity of their results. Surely, it helps to come up with ideas on the reality of viscous cosmology.

We now assume that the particle and the photon are positioned in a viscous surrounding. Then the total energy, (\ref{eq:tEnr1}), gets an additional contribution from the viscosity work, which apparently would slow down the expansion of the universe,
\bea \label{eq:tEnr2}
E&=&\frac{1}{2}m_p\, \dot a^2 - \frac{M\, m_p}{a} - \xi_p\, a^3\, \frac{\ddot a}{\dot a} + 2\pi - \frac{M\, m_{\gamma}}{a} - \xi_{\gamma} a^3 \frac{\ddot a}{\dot a},
\eea
where $\xi$ is the bulk viscosity coefficient. We assume that the expansion
of the universe is isotropic, that is, symmetric in all directions. Consequently, the shear viscosity coefficient likely vanishes. Comparing (\ref{eq:tEnr2}) with Friedmann's solution leads to another expression for the curvature parameter,
\bea \label{eq:k1}
k &=& - 2 \frac{E}{m_p} - 2 \frac{\xi_p}{m_p}\frac{\ddot a}{\dot a} a^3 + \frac{4 \pi}{m_p} - 2 \frac{\xi_{\gamma}}{m_p}\frac{\ddot a}{\dot a} a^3.
\eea

For the flat universe, $m_P$ has to be very large. Otherwise, when $m_p$ remains finite, the scale factor has to be given as follows
\bea \label{eq:k0}
a &=& \left(3\, \frac{2\pi -E}{\eta}\right)^{1/3}\; t^{1/3},
\eea
which apparently limits the total energy $E$ to be less than $2\,\pi$. For positively or negatively curved universes,
\bea \label{eq:kpm}
a &=& \left(3\, \frac{4\pi \mp m_p - 2 E}{2 \eta}\right)^{1/3}\; t^{1/3},
\eea
respectively. In deriving (\ref{eq:k0}) and (\ref{eq:kpm}), the approximation $\ddot a/\dot a\approx \dot a/a$ has been applied.
Otherwise, we have to find a solution for
\bea
\text{at } k = 0 && \hspace*{10mm} a^3\, \frac{\ddot a}{\dot a} =\frac{2\,\pi - E}{\xi}, \label{kkk0}\\
\text{at } k = \pm 1 && \hspace*{10mm} a^3\, \frac{\ddot a}{\dot a} = \frac{2\,\pi - E\pm m_p/2}{\xi}, \label{kkkpm}
\eea
To solve these equations, we introduce $\dot a =y$. Then, $\ddot a = y\, y_a^{'}$. Apparently, the solution of (\ref{kkkpm}), for instance, reads
\bea
a &=& \left(-\frac{3}{2}\, \frac{2\,\pi - E\pm m_p/2}{\xi} \right)^{1/3}\; t^{1/3},
\eea
which is almost compatible with (\ref{eq:kpm}).
When assuming that $E$ depends on $t$,
\bea
a &=& \left[-\frac{3}{2}\,\left(\frac{2 \pi}{\xi}\, t \pm \frac{m_p}{2 \xi}\, t - \int\frac{E(t)}{\xi}\, dt\;\right)\right]^{1/3}
\eea
Once again, the matter content defines the universe's curvature. The constraint on the total energy $E$ reads,
\bea
E &>& 2\, \pi \mp \frac{m_p}{2}.
\eea
So far, we conclude that the dependence of the scale factor $a$ on the co-moving time $t$ significantly changes with the viscous property of the background matter/radiation. It is illustrated in Fig. 1a, where an approximate comparison between ideal $a(t)\propto t^{1/2}$ and viscous background geometry $a(t)\propto t^{1/3}$ is illustrated. A comparison of the Hubble parameters in nonviscous and viscous background is given in Fig. 1b.

\begin{figure}[htb!]
\includegraphics[width=5.5cm,angle=-90]{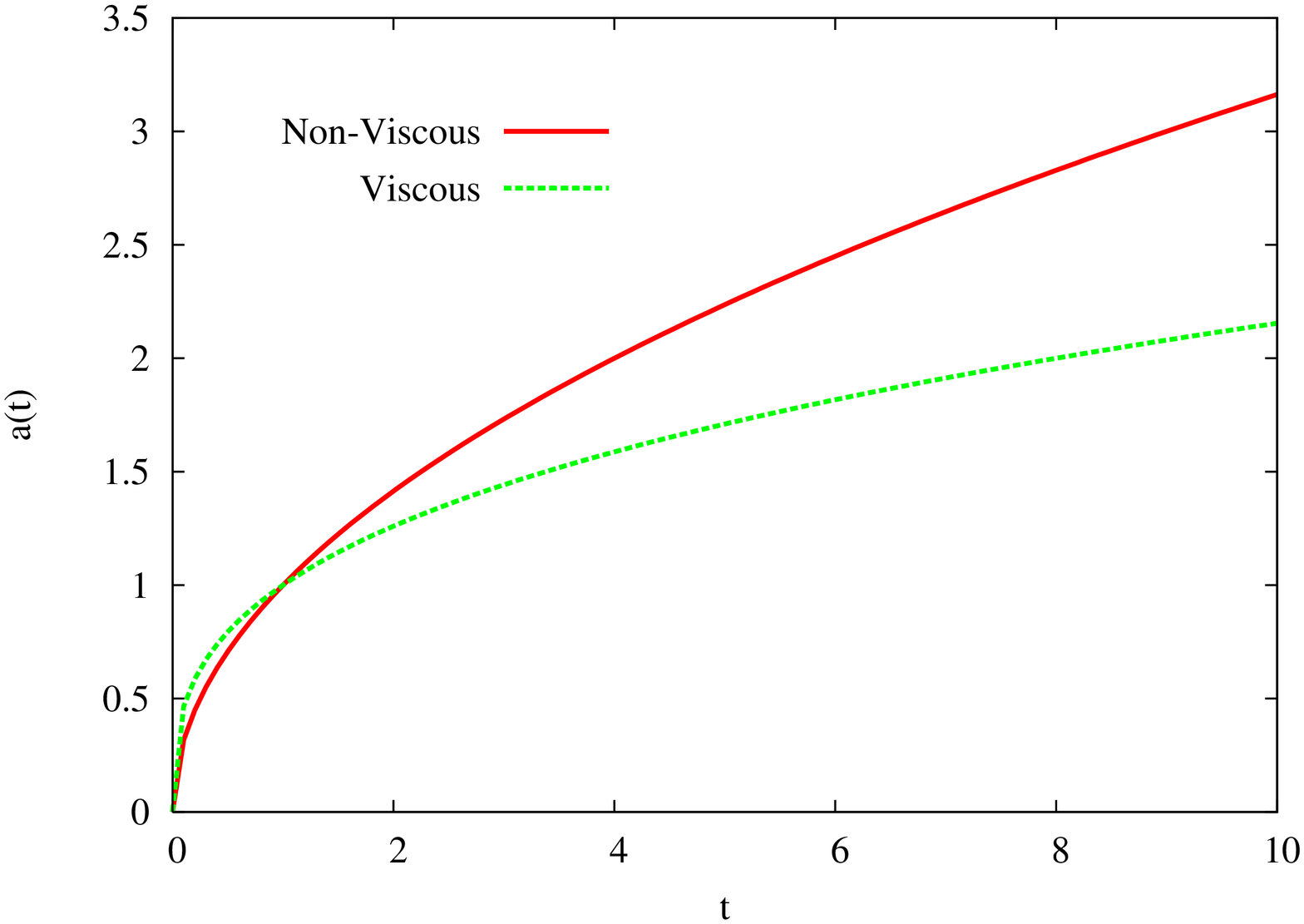}
\includegraphics[width=5.5cm,angle=-90]{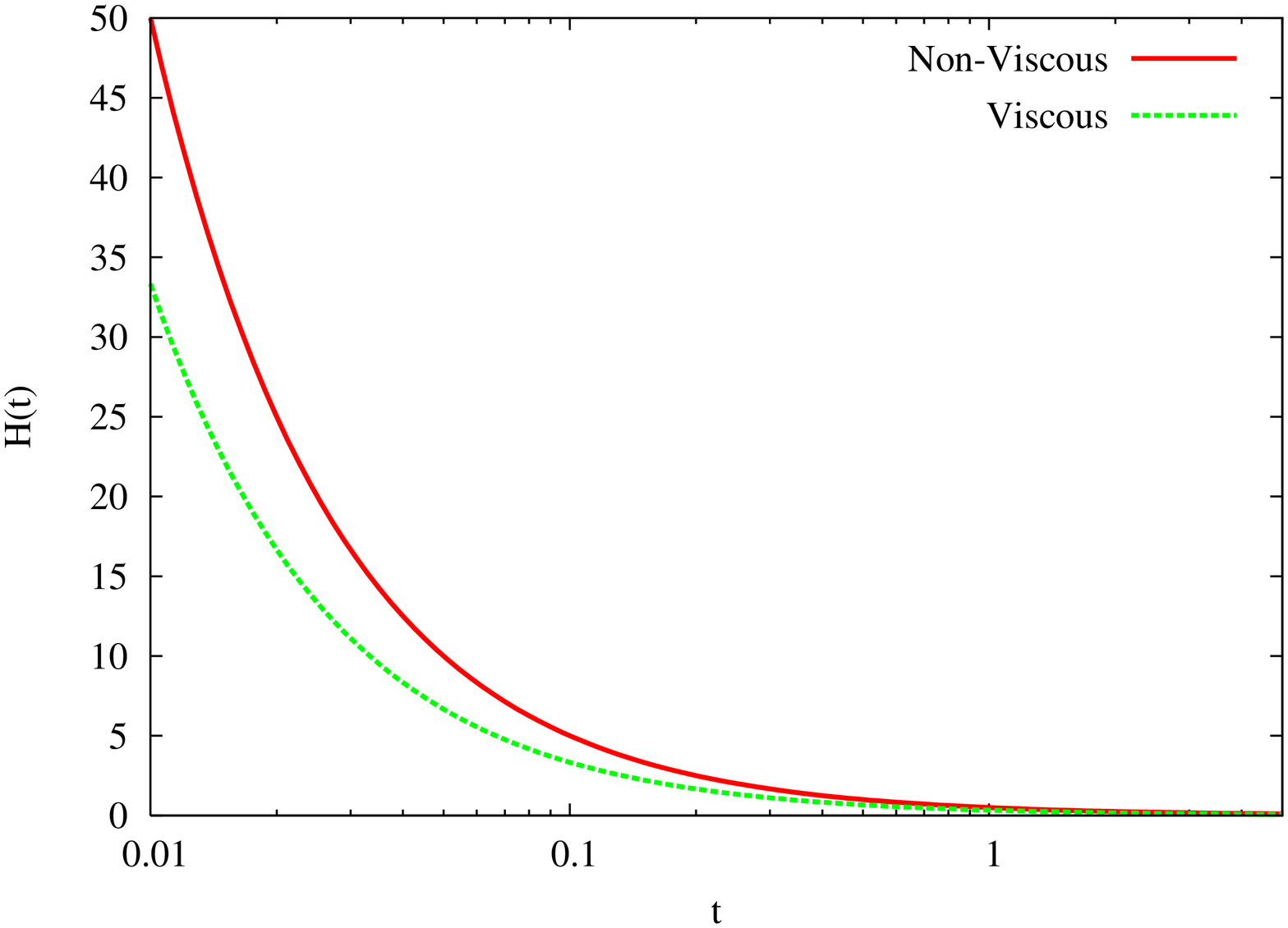}
\caption{(a)The dependence of the scale factor $a$ on the co-moving time $t$.(b)shows the Hubble parameter $H$ in dependence on $t$. Solid curve represents the case where the background geometry is filled with an ideal content. The dashed curve illustrates the effects of the viscosity. Because of the natural units, physical units are not given.}
\label{fig:1}
\end{figure}

\section{Rate of Energy Density in Viscous Cosmology} \label{sec:3b}

Based on the assumptions of the present model, it seems to be allowed to include the work of the bulk viscosity, Eq.~(\ref{eq:dq}). This results in
\bea \label{eg:du-eta}
dU &=& -\left(p\, dV - (\xi_p + \xi_\gamma)\, \frac{\ddot a}{\dot a}\, dV + \frac{2\pi}{a^3}\, dV\right).
\eea
Following the procedure given in section \ref{sec:2}, the last expression can be re-organized as done in (\ref{eq:adot1}). Then, the evolution of energy density, $\dot{\rho}$, does not   depend on the Hubble parameter $H$ only, but also on the thermodynamic ($\rho$ and $p$) and transport ($\xi$) quantities additional to,
\bea \label{eq:rhovis1}
\dot \rho &=& -3 \left((p+\rho) - \xi\,  \frac{d \dot a}{d a} + \frac{2\pi}{a^3} \right)\, H,
\eea
where $\xi=\xi_p+\xi_{\gamma}$. Comparing this evolution equation with the one in Eckart's relativistic fluid \cite{Eck40}, the  relativistic cosmic fluid leads to a direct estimation for the bulk viscous stress $\Pi$. The {\it ''conservation of total energy density''} or its static property requires that the bulk viscous stress equals the work of bulk viscosity.
\bea
\Pi&=&-\xi \frac{d\dot a}{d a}.
\eea
The total thermodynamic pressure, $P$, is given by summing up thermodynamic and viscous pressures
\bea
P &=& p + \frac{2\pi}{a^3}.
\eea
Obviously, this is one of the novel results of this model.

To estimate the evolution of the bulk viscous pressure, we adopt the causal evolution equation satisfying the $H$-theorem, non-negative entropy production, $S_{;i}^{i}=\Pi^{2}/\xi T\geq0$. Assuming that the total content in the background geometry is conserved, $T^j_{i; j}=0$, then the rate of the energy density has to fulfil the following conservation law:
\bea
\dot \rho &=& - 3\, H\, \left(p_{eff} + \rho\right),
\eea
where $p_{eff}=P+\Pi$. The comparison of this expression with (\ref{eq:drot}) illustrates the essential effect of the bulk viscosity, qualitatively. In order to make a quantitative comparison, we need to implement barotropic EoS, including one for $\Pi$.

%
%

\section{On the quantum cosmology} \label{sec:4}

In the previous sections, we have studied the universe as a closed system consisting of one particle, one photon and the {\it absolute} background. We have shown that this model is able to reproduce various results as the standard cosmological model. The quantum nature of such a system is still to be elaborated. Before doing this, some constrains have to be taken into account. First, we are far away from the quantization of the gravitational force \cite{gf1}. Second, the gravitational constant $G$ is taken as a universal constant, that is, it is valid always and everywhere \cite{gf2}. All proposals and even observations about time varying $G$ \cite{gtime1} are disregarded. Third, all ideas about the modification of the newtonian dynamics are not implemented \cite{mond}.

So far, we have checked  the case of one particle and one photon in both nonviscous and viscous surroundings. In the following, we suggest a quantum treatment. On the one hand, it is another check for the productivity and projectivity of the presented model. To this destination, we assume that the background geometry has $N$ particles and photons. These quantum particles are adhered within a cubic or spherical volume, $a^3$. Globally, the particles and photons are distributed, isotropically and homogeneously. Locally, the particles are distributed according to an occupation function, which depends on the particle's quantum numbers and correlations. The photons are obeying Bose–Einstein statistics. According to the standard cosmological model, the particles and photons are allowed to expand in a homogeneous and isotropic way. Then, the energy of a single particle in natural unites $E=(k^2+m^2)^{1/2}$, where the momentum $\vec{k}=(n_1\hat{x}+n_2\hat{y}+n_3\hat{z})2\pi/a$. To account for the interaction, we insert the potential $\phi$ and the correction of Uhlenbeck and Gropper \cite{uhlen1}.
\bea \label{eq:uhlen}
\int_0^a \exp\left(-\frac{\phi}{T}\right) \left[1\pm \exp\left(-m T r^2\right)\right] dr,
\eea
where $\pm$ refers to fermions and bosons, respectively. Then, the state density in the momentum space $a^3/h^3=V/(2\pi)^3$, where $a$ depends on $t$. The volume $V$ varies with $t$. Based on the proposed model, the volume of the universe can be determined by the size that is occupied by $N$ particles and photons,
\bea \label{eq:inta}
a(t)^3 &=&
       4 \pi^2\, N\, T(t)
       \left[ g_p e^{-\frac{\phi}{T(t)}}
       \left(2\, r \pm \sqrt{\frac{\pi}{m_p\,T(t)}}\right)
       \int_0^{\infty} \frac{k^2\,dk}{e^{\frac{E_p-\mu}{T(t)}}\pm 1} + \right. \nonumber \\
       && \hspace{22mm} \left.
	g_{\gamma} e^{-\frac{\phi}{T(t)}}
       \left(2\, r - \sqrt{\frac{\pi}{m_\gamma\,T(t)}}\right)
       \int_0^{\infty} \frac{k^2\,dk}{e^{\frac{E_\gamma-\mu}{T(t)}} - 1}\right]^{-1}, \hspace*{5mm}
\eea
where $r$ defines the region of interaction, $\mu$ is the chemical potential, and $g_p(g_{\gamma})$ is the degeneracy factor of the particle (photon). The potential $\phi$ will be introduced in section \ref{sec:sub1}.

It is apparent that differentiation with respect to the comoving time and dividing both sides by the scale factor $a$ results in the Hubble parameter
\bea \label{eq:Hhh}
H(t) &=&
  \left[\frac{\sqrt{m_p}}{12} \left(\sqrt{\pi}-2\, r \sqrt{m_g T(t)}\right)
       g_{\gamma} \int_0^{\infty} \textbf{csch}\left(\frac{\mu -E_{\gamma}}{2 T(t)}\right)^2 (\mu -E_{\gamma})\,k^2 dk\right] \frac{e^{-\frac{\mu}{T(t)}} \, \frac{dT(t)}{dt} }{A(t)} \nonumber \\
&-& \left[\frac{\sqrt{m_{\gamma}}}{3} \left(2\, r \sqrt{m_p T(t)} \pm \sqrt{\pi} \right) g_p \int_0^{\infty} \frac{e^{\frac{\mu+E_p}{T(t)}} (\mu -E_p)}{\left(e^{\frac{E_p}{T(t)}}\pm e^{\frac{\mu}{T(t)}}\right)^2}\, k^2\, dk \right]
\frac{e^{-\frac{\mu}{T(t)}} \, \frac{dT(t)}{dt} }{A(t)} \nonumber \\
&-&  \left\{\frac{\sqrt{m_{\gamma}}}{6} \left[4\, r \sqrt{m_p T(t)} (\phi - T(t)) \pm\sqrt{\pi} (2\phi - 3 T(t)) \right] g_p
    \int_0^{\infty} \frac{k^2 dk}{e^{\frac{E_p}{T(t)}}\pm e^{\frac{\mu}{T(t)}}}\right\} \frac{\frac{dT(t)}{dt}}{A(t)} \nonumber \\
&-& \left\{\frac{\sqrt{m_p}}{6} \left[\sqrt{\pi}\left(3 T(t)-2\phi\right) + 4\, r \sqrt{m_{\gamma} T(t)}(\phi - T(t))\right] g_{\gamma} \int_0^{\infty} \frac{k^2 dk}{e^{\frac{E_{\gamma}}{T(t)}}-e^{\frac{\mu}{T(t)}}} \right\} \frac{\frac{dT(t)}{dt}}{A(t)},
\eea
where
\bea
A(t) &=& \left[\sqrt{m_{\gamma}} \left(2 a \sqrt{m_p T(t)} \pm \sqrt{\pi}\right) g_p \int_0^{\infty} \frac{k^2 dk}{e^{\frac{E_p}{T(t)}}\pm e^{\frac{\mu}{T(t)}}}  \right. \nonumber \\
&-& \hspace*{2mm} \left. \sqrt{m_p} \left(\sqrt{\pi}-2 a \sqrt{m_{\gamma} T(t)}\right) g_{\gamma} \int_0^{\infty} \frac{k^2 dk}{e^{\frac{E_{\gamma}}{T(t)}}-e^{\frac{\mu}{T(t)}}} \right] T(t)^2. \hspace*{10mm}
\eea
It is essential to model $dT(t)/dt$ (section \ref{sec:sub2}) in order to have a numerical estimation for the time evaluation of $a(t)$ and $H(t)$. The hyperbolic trigonometric  function $\mathbf{csch}(x)=2e^x/(e^{2x}-1)$ diverges, when $x\rightarrow 0$, which is fulfilled in two cases; $T\rightarrow \infty$ or $\mu = E_{\gamma}$. The first condition is obvious, while $\mu = E_{\gamma}$ would mean that the chemical potential is very small; $\sim 2\pi$, where the photon`s energy and momentum are equal. The Expression (\ref{eq:Hhh}) is obviously the result of two contributions. The first two terms represent the contribution of the individual constituents as a free gas (collisionless). The correction of Uhlenbeck and Gropper, (\ref{eq:uhlen}), appears in the last two terms, in which the interactions between the individual constituents are taken into account through ${\cal O}(\phi)$,

\subsection{Photo–particle interactions}
\label{sec:sub1}

So far, we assume that the dynamics controlling the universe essentially originates in the gravitational interaction of the background with the particle  (matter) and with the photon (radiation), separately. At ultrahigh energy, the basic interactions between photon and matter appear in different types. Compton scattering describes an elastic interaction. In natural units, the photon`s energy can be given as  $2\pi/\lambda$, where $\lambda$ is the wavelength.
\bea
\Delta E_{\gamma} &=& m_p (1-\cos\langle\theta\rangle)^{-1}.
\eea
The scattering angle can be averaged as $\langle\theta\rangle\approx90^{o}$. Then, the Compton scattering potential $\phi_{cs}\approx m_p$. Summing up the gravitational potential with this value results in the total potential,
\bea \label{eq:phi1}
\phi &=& m_p\, + \frac{m_p m_{\gamma}}{a},
\eea
where $m_{\gamma}$ is the relativistic mass of the photon. When neglecting the relativistic mass of the photon, then Compton scattering results in a very small energy loss \cite{stecker1},
\bea
\frac{dE}{dt} &=& -\rho_{\gamma} \left(\frac{E}{m_p}\right)^2 = \frac{8 \pi^3}{a^3 m_p^2},
\eea
where $\rho_{\gamma}$ is the photon's density. Then, the rate of energy density reads
\bea
\rho(t) &=&- \frac{8 \pi^3}{a^6 m_p^2} \, t.
\eea

The second type of interaction is pair production, where $A+\gamma\rightarrow A + e^++e^-$, where the particle is participating in with its field. It may receive part of the energy released. The energies of a photon pair are partly absorbed in creating two electrons, $m_e$.
\bea
2 \pi &=& 2 m_e + K.E,
\eea
where the lift hand side represents the photon's energy.
For simplicity, it is conjectured that the total potential is given by the sum of the gravitational potential and the energy released from pair production,
\bea \label{eq:phi2}
\phi &=& 2 \left(\pi - m_e + \frac{m_p m_e}{a}\right) + \frac{m_e^2}{a}.
\eea
In an astrophysical context, the energy loss through pair production \cite{blumenthal1}
\bea
\dot \rho &=& - 3 \alpha \sigma \frac{(m_e T)^2}{(2 \pi)^3} f\left(\frac{m_e m_p}{2 E T}\right),
\eea
where the function $f$ has been calculated in Ref. \cite{blumenthal1}. It is obvious that these two types of interactions have to be subtracted from the total energy, (\ref{eg:rho-eta3}), (\ref{eg:rho-eta4}) and (\ref{eg:rho-eta5}).

The third type of photo–particle interactions is photo–pion production, where $A+\gamma\rightarrow A + \pi^++\pi^-$ \cite{pp1}. The most convenient way to describe the interaction between a particle (most likely a proton, where $Z=A=1$) and a photon in an observer's frame of reference is the invariant total energy $E_t$ in the center of momentum frame of reference (CMF), which moves with Lorentz factor $\gamma_{CMF} \cong E_p/E_t \gg 1$. The photon's energy $E_{\gamma}^{'}$ is calculated in the proton`s rest frame of reference \cite{pp2}.
\bea
E_t &=& \left(m_p^2 + 2 m_p E_{\gamma}^{'}\right)^{1/2},
\eea
where $E_{\gamma}^{'}=\gamma_p E_{\gamma} (1–\beta_p \cos \theta)$ and $\gamma_p=(1-\beta_p)^{-1/2}$.
The total potential is given by,
\bea \label{eq:phi3}
\phi &=& \left(m_p^2 + 2 m_p E_{\gamma}^{'}\right)^{1/2} + 2 \frac{m_p m_{\pi}}{a}.
\eea
Expressions (\ref{eq:phi1}), (\ref{eq:phi2}) and (\ref{eq:phi3}) can be used  in (\ref{eq:inta}) for Compton scattering, pair production, and photo–pion production, respectively.

\subsection{The time evolution of the temperature in the early universe}
\label{sec:sub2}

Expression (\ref{eq:Hhh}) is an essential input to estimate the energy density. For flat universe,  Friedmann's solution, (\ref{eq:adot1}), gives
\bea
\rho &=& \frac{3}{8 \pi} H(t)^2.
\eea
As given in (\ref{eg:rho-eta5}), the comoving time $t$ is related to the energy density as follows:
\bea
t & = & 6 \sqrt{\pi} \rho^{-1/2} = 12\, \pi\, \sqrt{\frac{2}{3}}\, H^{-1}.
\eea
Then the time derivative of temperature can be deduced as follows.
\bea \label{dTdt}
\frac{dT}{dt} &=&- \frac{\rho^{3/2}}{3 \sqrt{\pi}} / \frac{d \rho}{dT},
\eea
 which again depends on $d \rho/dT$, the dependence of $\rho$ on the cosmic temperature $T$. Modelling $d \rho/dT$ and replacing (\ref{dTdt}) in  (\ref{eq:inta}) makes it possible to prepare for an estimation for essential cosmological parameters, for instance $H$ given in (\ref{eq:Hhh}), in this quantum treatment.

\section{Discussion}
\label{sec:5}

So far, we show that a classical and quantum treatment of the universe results in {\it almost} quantitatively the same results as the ones deduced from the standard cosmological model. The qualitative behaviour of essential cosmological parameters is produced. For instance, in the radiation-dominated phase, in which $p=\rho/3$,(\ref{eq:rhovis1}) can be solved in the co-moving time $t$. In doing this, we utilize the results obtained from the dependence of $a$ and $H$ on the co-moving time $t$. Then,\ref{eq:adot1}) turns to offer a substitution of $\rho$ in $t$.
\bea \label{eg:rho-eta3}
\rho(t) &=& \frac{1}{36\, \pi}\; t^{-2} + \frac{6 \pi}{q^3}\; t^{-1} + \xi \, \frac{q^3}{9}\; t{-1},
\eea
where $q$ is the proportionality coefficient of  $a\propto t^{1/3}$. This is given in (\ref{eq:k0}) or (\ref{eq:kpm}). Expressions (\ref{eq:drot}) and (\ref{eq:pg1}) are deduced, when the matter filling the background geometry is nonviscous. Under the same assumptions, the solutions of these two expressions, respectively, read
\bea
\rho(t) &=& \frac{1}{36\, \pi}\; t^{-2} + \frac{6 \pi}{q^3}\; t^{-1}, \label{eg:rho-eta4} \\
\rho(t) &=& \frac{1}{36\, \pi}\; t^{-2}. \label{eg:rho-eta5}
\eea

\begin{figure}[htb!]
\includegraphics[width=10.cm,angle=-90]{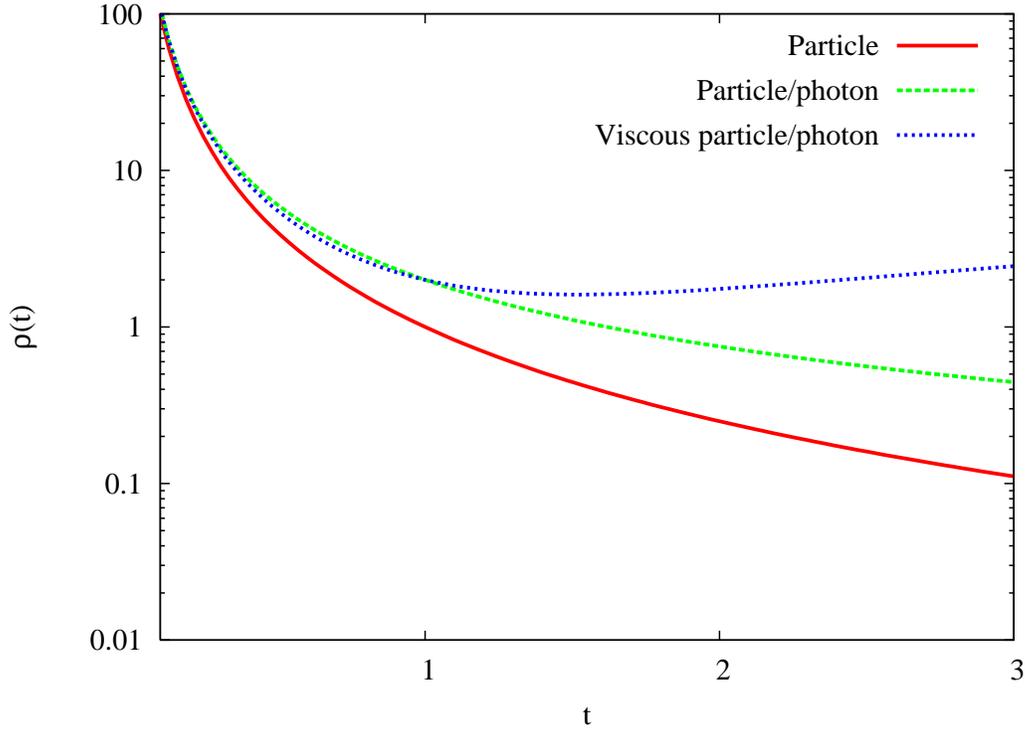}
\caption{An approximate comparison between the three cases given by (\ref{eg:rho-eta3}) and (\ref{eg:rho-eta4}). The solid curve represents the case where viscous matter and viscous radiation are filling the background geometry. The non-viscous contents are given by the dashed curve. Dotted curve stands for non-viscous background matter.}
\label{fig:2}
\end{figure}

Figure \ref{fig:2} shows an approximate comparison between the three cases given by (\ref{eg:rho-eta3}), (\ref{eg:rho-eta4}) and (\ref{eg:rho-eta5}). Seeking simplicity we disregard all coefficients, in other words, the proportionality $\rho \propto t^{\alpha}$ is merely drawn. This is not also valid for the fourth term in (\ref{eg:rho-eta3}), which has units of energy density. Therefore, no physical units can be deduced. It is obvious that the contribution to the energy density differs over the comoving time $t$. At a very early stage, the viscosity adds with a negligible amount to the energy density. Later on, we notice that the viscous contents seem to become dominant. It leads to a small increase in $\rho$ with increasing $t$. As it is included in (\ref{eg:rho-eta3}) with a positive sign, it apparently sets a limitation for the validity of the presented model. At the limit, where $\rho$ increases with increasing $t$, the model seems to case being valid.

A few comments are now in order. Along the entire history of this universe, we are assuming that the background geometry is filled with particles and photons. The dynamics controlling the cosmological evolution is determined by these two constituents, which are treated as nonviscous and viscous fluid. Furthermore, we assume that no phase transition is taken into consideration. Therefore, the presented model seems to assume that a certain phase remains unchanged, while the universe was expanding. The limitation of viscosity appearing in Fig. \ref{fig:2} likely would refer to the necessity of the phase transition accompanied by symmetry changing, for instance.

The time evolution of the energy density has been studied in different contents filling the background geometry. First, we start with matter, (\ref{eq:drot}). When adding photon (radiation), we get ({\ref{eq:pg1}). The effect of the viscosity appears in (\ref{eq:rhovis1}).

In section \ref{sec:4}, we make a step further towards the quantum treatment. We assume that the partition function of a {\it closed} system consisting of $N$ particles and photons is able to describe the entire universe. The size of the universe is given by the volume occupied by these two constituents. The interactions between particle and photon are taken into account. All cosmological parameters, like $a$, $H$  and $\rho$ can be deduced. Therefore, the expansion of the universe seems to be accessible.

As an outlook, we may want to check the effects of the phase transitions and the corresponding changes in the degrees of freedom and EoS. Also, we are planning to study the potential change when taking into account the relativistic mass of the photon.

\end{document}